\documentclass[aps,preprint,showpacs,superscriptaddress]{revtex4-1} 
\usepackage{amsmath,amssymb,graphicx,epstopdf}
\begin{document}

\title{Ultrabroadband dispersive radiation by spatiotemporal oscillation of multimode waves}

\author{Logan G. Wright}
\email{ lgw32@cornell.edu}
\affiliation{School of Applied and Engineering Physics,
Cornell University, Ithaca, New York 14853, USA}
\author{Stefan Wabnitz}
\affiliation{Dipartimento di Ingegneria dell$'$Informazione, Universit\`{a} degli Studi di Brescia, and Istituto Nazionale di Ottica, CNR, via Branze 38, 25123 Brescia, Italy}
\author{Demetrios N. Christodoulides}
\affiliation{CREOL, College of Optics and Photonics, University of Central Florida, Orlando, Florida 32816, USA}
\author{Frank W. Wise}
\affiliation{School of Applied and Engineering Physics,
Cornell University, Ithaca, New York 14853, USA}

\begin{abstract} Despite the abundance and importance of three-dimensional systems, relatively little progress has been made on spatiotemporal nonlinear optical waves compared to time-only or space-only systems. Here we study radiation emitted by three-dimensionally evolving nonlinear optical waves in multimode fiber. Spatiotemporal oscillations of solitons in the fiber generate multimode dispersive wave sidebands over an ultrabroadband spectral range. This work suggests routes to multipurpose sources of coherent electromagnetic waves, with unprecedented wavelength coverage. 
\end{abstract}


\maketitle 

Eigenmodes are ubiquitous tools for describing complex wave systems. For nonlinear complex wave systems, however, the superposition principle is not applicable. In special cases, nonlinear wave systems possess solitons, which act to some extent as nonlinear eigenmodes. Combined with more general nonlinear attractors and perhaps insights from linearized systems, researchers may thus build up a conceptual understanding of complex nonlinear wave dynamics. In optics, one-dimensional (1D) dynamics in single-mode waveguides have been thoroughly explored, with many advances hinging on the robust nonlinear attraction of solitons\cite{Hasegawa1973,Mollenauer1980,Kivshar2003}. In unbounded 3D systems, dynamics have been successfully explained largely in terms of a nonlinearly attracting instability, namely spatial or spatiotemporal collapse\cite{COUAIRON2007}. In reality, the collapse singularity is avoided by a multitude of higher-order effects, and the field eventually expands, by propagating linearly. Post-collapse, promising results have been obtained considering conical waves, which are the eigenmodes of the 3D linear wave equation\cite{Faccio2006,Majus2014}. Although not yet widely adopted, conical wave solutions to the nonlinear wave equation may provide even deeper insight\cite{Conti2004,Porras2007}. Given the great advantages the concepts of solitons and collapse have provided in the field of nonlinear optical waves for studying single-mode waveguides and free space filamentation, it is natural to seek similarly useful nonlinear objects in multimode waveguides. Multimode waveguides include as limiting cases single mode (1D) and free-space (3D), so these hypothetical objects, whether solitons, nonlinearly attracting instabilities, guided conical waves, or something else entirely, could even help to conceptually unify nonlinear optical dynamics across dimensions. 

Solitons in optical single-mode fiber (SMF) have been intensely researched, because they are relatively accessible both analytically and experimentally. Equally important, soliton dynamics are critical to telecommunications, mode-locked fiber lasers, and compact white-light sources with high spatial mode quality. Multimode fibers (MMFs) could provide major benefits for various applications, from spatial division multiplexing in communications\cite{Winzer2012,Richardson2013}, to high-power, versatile fiber lasers and white-light sources\cite{Wright2015b}. Although wave propagation in MMF is still experimentally and theoretically challenging, recent theoretical advances\cite{Poletti2008,Mafi2012,Andreasen2012,Mecozzi2012,Mecozzi2012a,Mumtaz2013} combined with opportunities for major performance increases provide ample motivation for their study. From a scientific perspective, MMF is an ideal environment for studying spatiotemporal nonlinear optical dynamics. By judicious design, or by control of the initial excitation, researchers may control the spatiotemporal characteristics of nonlinear dynamics, through variation of the effective dimensionality, the coupling between modes, or their individual dispersions.

These factors have motivated recent work on nonlinear optical waves in MMFs\cite{Mecozzi2012,Mecozzi2012a,Renninger2013,Russell2014,Tani2014,Buch2015,Wright2015a,Wright2015b}. In particular, control of the spatial excitation of a MMF provides a means of controlling the nonlinear spatiotemporal dynamics, and therefore the characteristics of a supercontinuum\cite{Wright2015b}. Launching $\approx$200-nJ and 500-fs pulses at 1550 nm into a graded-index (GRIN) MMF produces remarkable visible light emission characterized by a series of spectral peaks with non-uniform spacing. Numerical simulations confirm these peaks, and suggest that even more spectacular emission occurs at long wavelengths, where fused silica is effectively opaque (Figure 1). These observations are puzzling and fascinating: some novel nonlinear optical mechanism, unidentified by prior fiber-based or bulk supercontinuum research, generates and links electromagnetic radiation over 2 orders of magnitude in wavelength (from $>$ 50 to $<$ 0.5 $\mu$m). 

\begin{figure}[h]
\centering\includegraphics[width=12. cm]{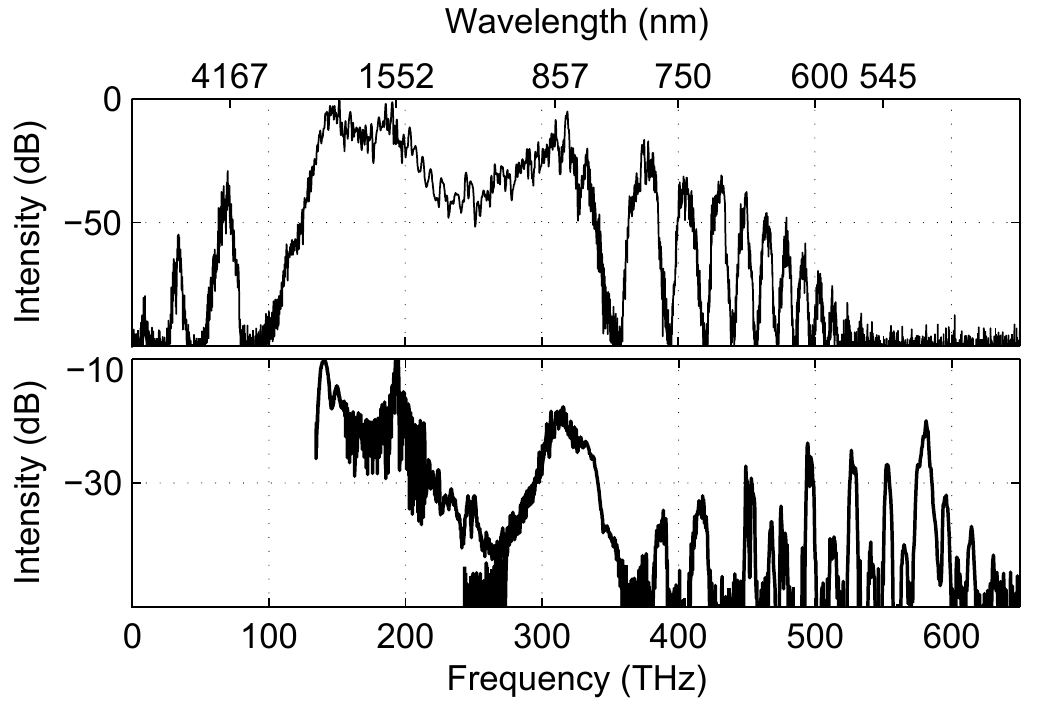}
\caption{Simulated (top) and example experimental (bottom) supercontinuum in multimode GRIN fiber. The pump pulse at 1550 nm creates a spectrum with a series of red-shifted and blue-shifted peaks. In the bottom panel, the y-axis reference (0 dB) is the maximum intensity of the 1550 nm pump peak.}
\end{figure}

Here we provide a theoretical explanation for this phenomenon: dispersive waves are generated by the spatiotemporal oscillation of multimode solitons. The process is inherently 3D, with spatiotemporally-evolving nonlinear waves emitting spatiotemporally-evolving dispersive waves. However, we show that insights from solitons of the 1D nonlinear Schr\"{o}dinger equation (NSE), can prove useful in understanding the higher-dimensional system.  This understanding also suggests routes to generating ultrashort pulses in fiber at wavelengths outside current capabilities, and may provide a means of interfacing normally distinct parts of the electromagnetic spectrum.

Solitons are solutions to a conservative equation. To use them in applications such as mode-locked lasers and optical telecommunications\cite{Mollenauer1984,Mollenauer1985,Haus1985,Mollenauer1986,Hasegawa1989,Smith1990,Ellis1991,Nakazawa1991,Richardson1991,Gordon1992,Matera1993,Haus1996}, loss must be compensated. This is naturally accomplished by laser gain, or by amplifiers placed throughout the line. The periodic perturbations of gain and loss, however, can destabilize a soliton. The origin of the instability, and the characteristic spectral sidebands that are its signature, is the fact that a periodic perturbation can phase-match dispersive wave emission at particular frequencies\cite{Pandit1992, Kelly1992}. For a perturbation period (spacing of amplifiers, or laser cavity length) $Z_c$, the phase-matching condition is approximately:
\begin{equation}
(k_{sol}-k_{dis})=2m\pi/Z_c
\end{equation}
where $m$ is an integer, and $k_{dis}$ is the wavevector of the dispersive wave. $k_{sol}$ is the soliton wavevector, which in the absence of higher-order dispersion is equal to $|\beta_2|/{2\tau^2}=\pi/(4Z_o)$,where $\beta_2$ is the group velocity dispersion, $\tau$ is the soliton duration and $Z_o=\pi/(4k_{sol})$ is the soliton period. This quasi-phase matching leads to resonant emission at frequencies separated from the pump by
\begin{equation}
\Omega_{res}=\frac{1}{\tau}\sqrt{\frac{8Z_om}{Z_c}-1}
\end{equation}
\cite{Gordon1992,Kelly1992,Mollenauer1989}. This result was later refined to include the third-order dispersion of the fiber\cite{Dennis1993,Dennis1994,Kodama1994}, where resonant frequencies were found to be roots of the equation:
\begin{equation}
(k_{sol}-k_{dis})=2m\pi/Z_c=(\frac{1}{2}+4b_3^2-8b_3^4)-(-\frac{\Omega^2}{2}-b_3 \Omega^3)
\end{equation}
where $b_3 = -\beta_3/(6|\beta_2|\tau)$ ($\beta_3$ is the third-order dispersion) and $\Omega$ is the angular frequency separation from the pump (in units of $\tau^{-1}$). Production of dispersive waves is the primary limitation to the performance of soliton fiber lasers.

Although they were predicted as early as the 1980s\cite{Hasegawa1980,Crosignani1981,Crosignani1982,Yu1995,Raghavan2000}, solitons consisting of pulses within multiple spatial modes have only recently been studied experimentally\cite{Renninger2013,Wright2015a,Wright2015b}. Initial work shows that solitary waves, termed MM solitons, can be excited in multimode graded-index (GRIN) fibers. Similarly to single-mode solitons, these pulses resist group-velocity dispersion and adjust themselves adiabatically in response to perturbations such as intrapulse Raman scattering. This behavior makes them a useful conceptual tool for understanding complex nonlinear processes beyond the NSE description, such as those involved in supercontinuum generation\cite{Wright2015a,Wright2015b}. Unlike single-mode solitons however, MM solitons additionally resist modal velocity dispersion - \textit{i.e.}, the fact that each of the numerous consitutent modes of the soliton has a different group velocity. Fission of MM solitons is spatiotemporal: it yields multiple MM solitons which can have many different modal distributions. In fact, MM solitons can have a bewildering range of spatiotemporal shapes, which makes their dynamics much richer and challenging than single-mode solitons. While they are in some sense natural extensions of the 1D NSE soliton to the case of a MM fiber, they probably do not fulfill the most rigorous definitions of $'$soliton$'$ (some special cases are already known, however\cite{Mecozzi2012,Mecozzi2012a}). Questions such as whether they are stable over very long distances, how they interact with one another, and how many modes can be involved, remain largely unanswered. 

When a beam excites multiple modes of a GRIN fiber, it propagates through the fiber with a characteristic spatial oscillation with period (pitch) $P = \pi R/\sqrt{2\Delta}$, where $R$ is the core radius and $\Delta$ is the core-cladding index difference\cite{Renninger2013,Buck2004}. This oscillation causes the intensity of the beam to periodically evolve, as in a loss-managed soliton transmission line or soliton fiber laser. For a pulsed beam, oscillations will only occur as long as the pulses in each mode maintain co-localization and a phase relationship\cite{Renninger2013}. Hence, a soliton containing multiple spatial modes experiences a periodic oscillation of its peak intensity and therefore is likely to emit dispersive radiation at particular frequencies depending on the period of the oscillation. 

\begin{figure}[h]
\centering\includegraphics[width=8. cm]{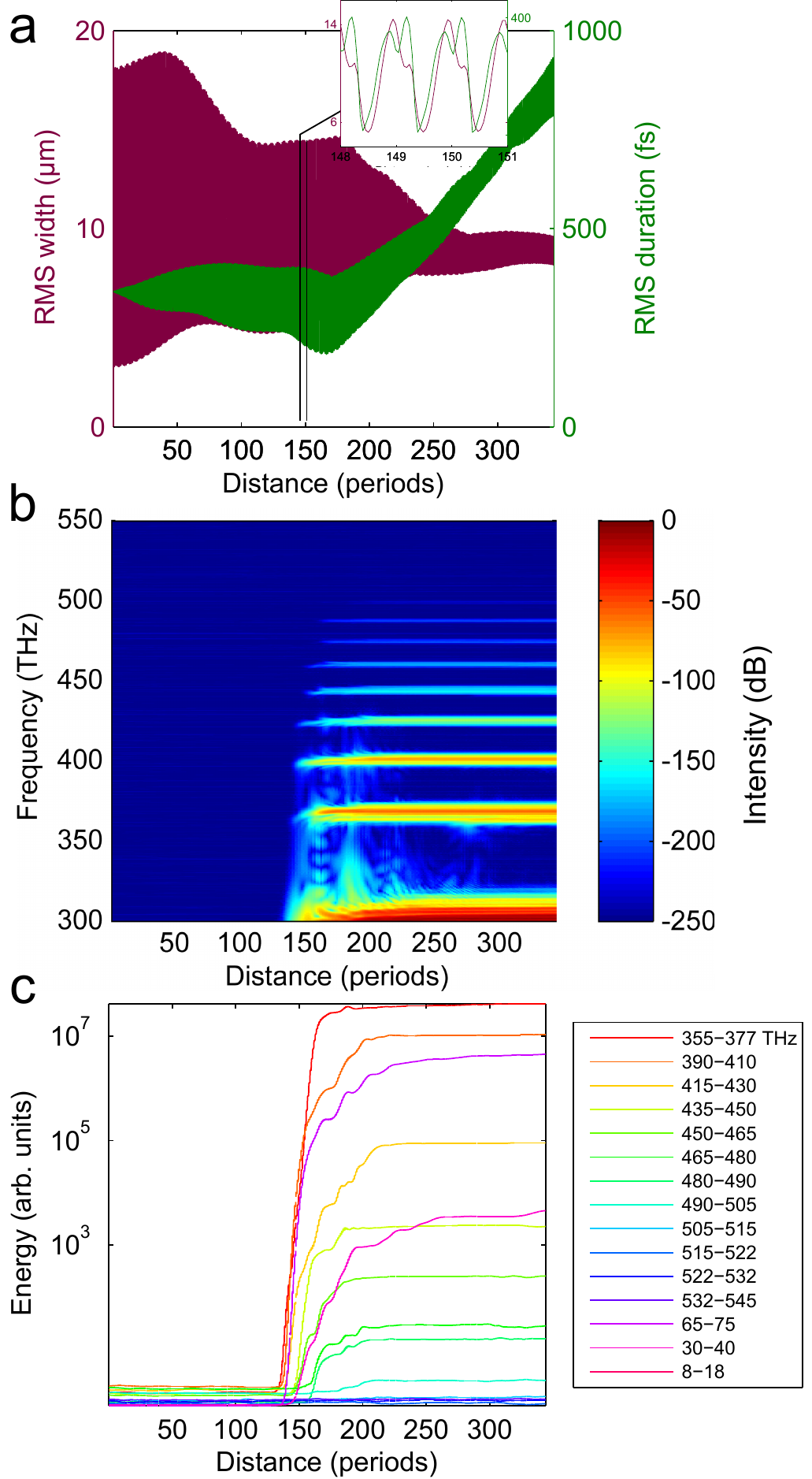}
\caption{Dynamics of dispersive wave formation in GRIN fiber. a) Temporal and spatial breathing of the field (inset: zoom in near the onset of dispersive wave generation). b) Evolution of the spectral intensity of the whole field through the same distance. c) Energy in each dispersive wave band. The x-axis scales are normalized to the linear spatial oscillation period of the GRIN fiber, equal to 407 $\mu$m. These dynamics are also shown in Supplementary Movies 1 and 2, which provide a considerably more complete representation of the complex spatiotemporal evolution.}
\end{figure}
\begin{figure}[htb]
\centering\includegraphics[width=9. cm]{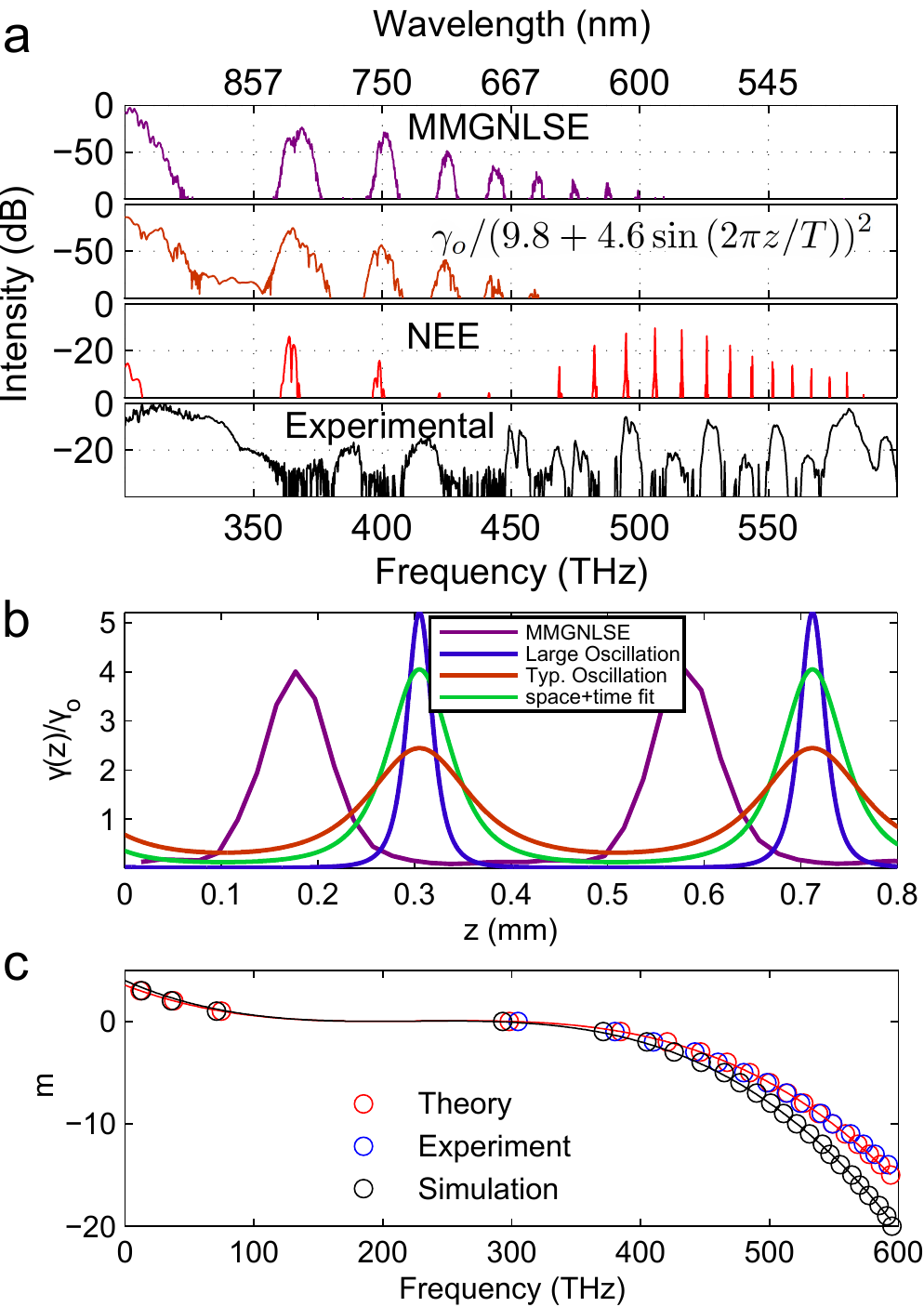}
\caption{a. ODW from MMGNLSE simulations (MMGNLSE), eqn. 4 with $\gamma(z)$ as in text,1D NEE simulations, and an experimental example spectrum. b. Functions used in the quasi-1D approximation, along with the peak intensity variation of the field in MMGNLSE simulation. c.Comparison of the periodic 1D phase-matching model with multimode simulations by the GMMNLSE and experiments in GRIN MMFs. Continuous curves are values of $m$ in Eq. 3, plotted with the best-fit parameter values.}
\end{figure}

We consider a simulation using the generalized multimode nonlinear Schr\"{o}dinger equation (GMMNLSE)\cite{Poletti2008}, with the first five radially-symmetric modes excited uniformly for simplicity (Figure 1). Experiments are conducted with the procedures described in Ref. \cite{Wright2015b}. In both simulation and experiment, we find that the soliton oscillation-induced dispersive waves (ODWs) are observed for many different initial spatial conditions, including non-radially-symmetric excitations. Typically we observe that the ODWs are more apparent when the intensity of the initial pulse is higher (either by large energy, tight spatial localization, or both). Changing the initial noise level or the pulse duration results in different dynamics. Nonetheless, provided the pulse energy and fiber length are sufficient, and that multiple modes are excited, the qualitative features (including the spectral positions) of the ODW emission process are similar. In simulations, we observe that the energy of each ODW is distributed roughly equally among all the modes, with the red-shifted (blue-shifted) sidebands exhibiting a slight preference for the low-order (higher-order) modes. 

Representative simulation results for a 400-fs, 1550-nm MM pulse with energy 164 nJ launched into a GRIN MMF are shown in Figure 2. (The results are also shown as Supplementary Movies 1 and 2.) The spatial and temporal breathing of the field are evident in Fig. 2a. While we show only the blue-shifted ODWs (Figure 2b), the simulation also predicts red-shifted ODWs. These are remarkably outside the transparency window of fused silica (the first appears at roughly 72 THz (4200 nm)). As the pulse traverses the fiber, the spectrum develops a dispersive wave near 300 THz (1000 nm), as well as ODWs in the visible and mid-IR regimes. Attenuation is included in the simulations with an assumed frequency dependence $\alpha=\alpha_{1550}\exp{-(f-c/(1550 nm))/b_l}$, where $\alpha_{1550}= 0.05 dB/km$ is the attenuation (units m$^{-1}$) at 1550 nm, and $b_l = 0.0062$ PHz ($\approx$ 80 nm) models the increasing loss into the infrared. For these parameters, attenuation is $\approx$ 0.1 MdB/km at 4200 nm. We neglect loss above the 1550-nm base level on the blue side of the spectrum. Despite this enormous loss, appreciable energy is still produced in the mid-IR  ODWs. This is partly due to the relatively short length of the fiber (150 periods correspond to only $\approx$ 6 cm), and to the enormous gain the ODWs experience (relative to the initial continuum level). The energy in each ODW may grow superexponentially, with the first blueshifted (redshifted) ODW experiencing roughly 70 dB (50 dB) net gain in 1 cm (Figure 2c). After about 100 oscillation periods the ODW energy in each spectral band saturates to a nearly constant value. The observed superexponential growth and subsequent saturation of the ODW energy is in qualitative agreement with the analytical prediction of the soliton perturbation theory that was developed to describe large-amplitude soliton intensity oscillations in periodically amplified transmission links\cite{Georges1996,Midrio1996}. On the other hand, for small amplitude oscillations of the soliton, the theory only predicts a linear growth of the ODW with distance\cite{Elgin1993}. 

To verify that the oscillations underly the ODWs, we add oscillations artificially to a single-mode NSE. This is done by making the nonlinear coefficient a periodic function of the longitudinal coordinate, so that the pulse evolution equation is

\begin{equation}
\frac{\partial A(z,t)}{\partial z}=-i\frac{\beta_{2}}{2}\frac{\partial^{2} A(z,t)}{\partial t^{2}}+\frac{\beta_{3}}{6}\frac{\partial^{3} A(z,t)}{\partial t^{3}}+ i\gamma(z)|A(z,t)|^{2}A(z,t) 
\end{equation}
where $A(z,t)$ is the pulse envelope, and $\gamma(z)$ is the $z$-dependent nonlinear coefficient. Figure 3 compares the result of solving this equation to the results above found using the GMMNLSE, and with experiment. Figure 3a shows the solution of equation 4 with the indicated form of $\gamma(z)$. Due to the sinusoidal oscillation of the beam radius, one expects an accurate $\gamma(z)$ to be of the form $\gamma(z) =  \gamma_o/([r_{max}-r_{mean}]+r_{mean}\sin{(2\pi z/P)})^2$. We use the RMS widths for $r_{max}$ and $r_{mean}$ here, and rescale $\gamma_o$ appropriately, since the usual spatial width measurement (mode field diameter) is not well-defined for the complex spatiotemporal fields that occur. At the onset of the ODW generation ($\approx$ 145-155 periods), $r_{max}=$ 14.5 $\mu$m and $r_{min}=$ 5.2 $\mu$m, so that $\gamma(z) = \gamma_o/(9.8+4.6\sin{(2\pi z/P)})^2$ (1Db, Fig. 3c). The observed intensity oscillations of the MMF field (MMGNLSE, Fig. 3b) are approximated better by $\gamma_o/(9.8+6.9\sin{(2\pi z/P)})^2$ (space$+$time fit, Fig. 3b). This is because the MM soliton's duration also oscillates: the soliton tends to lengthen when its spatial width increases and shorten when its spatial width decreases (Fig. 2a and inset, Supplementary Movie 1). It is this spatiotemporal oscillation that generates the ODWs.

The experimental peak locations are consistent with simulation and analytic theory. Figure 3c shows the results of fitting the experimentally-measured and simulated peak locations with the roots of Eq. 3 for various $m$. For the experimental (simulated) peaks, fitting yields $\beta_2$ = -26 (-25) fs$^2$/mm, and $\beta_3$=143 (143) fs$^3$/mm. In both cases, the peaks are fit by the simple theoretical approximation well, with the discrepancy attributed to the slightly different dispersion characteristics of the real and simulated fibers. The optimal values are near the simulation parameters, except that the optimal $\beta_2$ is greater by 16\%, which suggests that modal dispersion makes an effective contribution. 

Although the measured ODW positions are well-described, their amplitudes can vary due to several effects beyond both the simulations and analytic approach. First, due to the presence of many modes in the fiber, phase-matched intramodal four-wave mixing (FWM) may occur involving the 1550-nm pump, its primary dispersive wave at 800-1000 nm, various ODWs, and Raman-shifted MM solitons. For the low-order ODWs in particular, we note that there are several energy-conserving FWM processes. These may amplify red-shifted ODWs at the expense of specific blue-shifted ODWs. Particularly in the presence of many modes, intramodal FWM can be a complex process\cite{Guasoni2015}, particularly when cascaded mixing is considered\cite{Demas2015}. It is probably why, for certain initial spatial conditions, various low-order blueshifted ODWs are attenuated or even invisible experimentally. We obtained the peak locations in Figure 3c from experiments with multiple initial conditions, in order to account for low-order blue-shifted ODWs that were attenuated for any given intial condition. Second, because the spacing of the ODWs is quite close to the Raman bandwidth of fused silica, certain ODWs may experience Raman gain from one another and from the third-harmonic light (THG: its efficiency may also benefit from intramodal phase-matching). Lastly, the dynamic range and spectral resolution of the spectrometer limits the visibility of low-amplitude features, and broadens narrow spectral features.

Another remaining mystery is the relatively high amplitude of the ODWs observed in experiments, compared to simulations. Intramodal FWM may play some role, as well as larger oscillations and THG. Figure 3a shows the spectrum obtained by solving the 1D nonlinear envelope equation (NEE) with $\gamma(z) = \gamma_o/(15.5+14.5\sin{(2\pi z/P)})^2$, including self-steepening and assuming averaging of the oscillating Raman integral\cite{Genty2007}. Larger oscillations produce relatively more intense dispersive waves, because the soliton is more strongly perturbed. Larger oscillations occur when more modes are coherently locked together\cite{Wright2015a}, and as the experiment contains much more than 5 modes, we choose a functional form of $\gamma(z)$ to model this. In addition, THG provides an initial seed in the visible, leading to significantly enhanced resonant emission at nearby wavelengths. Periodic backconversion occurs for frequencies near $3\omega_o$, so the largest enhancement is for slightly smaller frequencies. 

The ODWs generated by soliton resonances in MMF have relevance to applications. For example, by filtering the first redshifted or blueshifted ODW, one may generate pulses in wavelength regions well outside the gain spectrum of available fiber dopants.  Tuning may be achieved by changing the pump wavelength or fiber pitch. In fact, because the modulation instability gain spectrum of a CW field at the pump wavelength overlaps with the soliton sidebands\cite{Matera1993}, filtered ODWs could be parametrically amplified (using either the circulating pump pulse or an injected CW field) provided some mechanism was introduced to compensate for the chromatic walk-off between the pump and ODW fields. ODWs spaced over a very broad wavelength range are generated coherently from the optical pump pulse. Given the remarkable range and coherence, spatiotemporal soliton oscillation may provide a means of generating synchronized ultrashort pulses in different regions of the electromagnetic spectrum. In an appropriate waveguide, an optically-derived coherent connection between microwave, deep ultraviolet and optical frequencies may be possible. 

In summary, we have shown that the spatiotemporal oscillation of nonlinear waves in GRIN multimode fiber causes the generation of their spatiotemporal dispersive waves. These dispersive waves can be described relatively well by simulations using the GMMNLSE, and insight can be gained by approximating the dynamics in a quasi-1D model with a longitudinally-varying nonlinearity. Future work, involving more advanced models and experimental methods, can answer a few of the open mysteries about this process, including verification of these hypotheses, the spatiotemporal dynamics leading to the multimode supercontinuum, and the spatiotemporal structure of the dispersive waves. This work illustrates valuable conceptual connections between 1D and the intermediate dimensional system of the GRIN MMF. Practically, it provides a route to fiber-based ultrashort pulse sources with tunable wavelengths far outside the range of any current fiber-optic technique.

\section*{References}

\end{document}